# Observation of Fully Flat Bands in a Photonic Dipolar Kagome Lattice

Han-Rong Xia,[1,2] Ziyao Wang,[2] Yunrui Wang,[1] Zhen Gao,[2,*] Meng Xiao[1,3,*]

[1] Key Laboratory of Artificial Micro- and Nano-structures of Ministry of Education and School of Physics and Technology, Wuhan University, Wuhan 430072, China

[2] State Key Laboratory of Optical Fiber and Cable Manufacture Technology, Department of Electronic and Electrical Engineering, Southern University of Science and Technology, Shenzhen 518055, China

[3] Wuhan Institute of Quantum Technology, Wuhan 430206, China

Corresponding Email [*]: gaoz@sustech.edu.cn; phmxiao@whu.edu.cn

**Abstract**

Flat bands, characterized by zero group velocity and strong energy localization, enable interaction-enhanced phenomena across both quantum and classical systems. Existing photonic flat-band implementations were limited to evanescent-wave systems, specific lattice symmetries, or complex supercell modulations. A simple, universal, and efficient approach to realizing flat bands without dedicated source excitation is to be explored. Here, inspired by geometrically frustrated configurations, we theoretically proposed and experimentally demonstrated threefold-degenerate flat bands by integrating orbital and rotational degrees of freedom in a photonic dipolar kagome lattice. By rotating the dipole orientation, the system exhibits a band flip transition at which point all bands achieve complete flatness and degeneracy across the entire Brillouin zone. In contrast to conventional s-orbital kagome lattices with only a single flat band, our approach flattens the entire band structure, eliminating dispersive modes and enabling compatibility with arbitrary excitations. These results establish a new mechanism for flat-band engineering, offering a tunable strategy for enhancing light-matter interactions and may have applications in compact photonic devices and energy-efficient information processing.

**I. Introduction**

In periodic systems, particles such as electrons and photons obey Bloch's theorem, forming extended Bloch waves with dispersive energy bands. However, when the energy becomes independent of the wave vector, the system exhibits a flat band, characterized by vanishing group velocity and highly localized eigenstates [1,2]. This suppression of kinetic energy enhances the effects of interactions and geometric constraints, enabling a wide range of unconventional physical phenomena across both quantum and classical systems [3-16]. In condensed matter physics, flat bands have been proposed as an ideal platform for realizing strongly correlated phases such as Mott insulators [3-5], fractional Chern insulators [6,7], and unconventional superconductivity [8-10]. In particular, twisted two-dimensional (2D) materials with moiré superlattices near the magic angle host ultra-flat bands that generate correlated insulating states, superconductivity, orbital magnetism, and strongly interacting Hubbard-like physics [3-5,8]. In classical systems such as photonics, acoustics, and mechanics, flat bands have enabled non-diffractive beam propagation [11,12], enhanced light-matter interactions [13], robust mode confinement [14], and low-threshold lasing [15,16], offering practical routes toward compact devices.

Generally, flat bands can be realized through various mechanisms, including external magnetic fields [17,18], synthetic gauge fields [19-22], engineered mode coupling [23], twist engineering [24-26], Floquet modulations [27-29], and synthetic dimensions [30,31]. A particularly straightforward and experimentally accessible approach, especially in 2D photonic systems, involves lattice geometry and symmetry design [11,12,32-39]. Prototypical examples include Lieb and kagome lattices that utilize destructive interference to suppress dispersion [11,12,38]. However, the flat bands in these systems are typically degenerate with dispersive bands at high-symmetry points, making the selective excitation of flat bands challenging and weakening the clarity of flat-band-associated phenomena. Moreover, realistic photonic systems often involve long-range or higher-order couplings—factors commonly neglected in these ideal lattice models, which will reintroduce dispersion and destroy the flat bands.

In this work, inspired by geometric frustration—a condition where the lattice geometry prevents the simultaneous satisfaction of local constraints, we theoretically propose and experimentally demonstrate a new flat-band mechanism in photonic crystals (PhCs). We construct a dipolar kagome lattice featuring *p*-orbital modes with tunable dipole orientations and realize three fully flat and energetically isolated bands. This configuration supports robust mode localization under arbitrary excitations, exhibits a band flip transition controlled by dipole rotation, and remains resilient to higher-order couplings which are usually unavoidable in PhCs. Our findings uncover a simple, tunable, and general strategy for engineering flat bands, which might have important applications in enhancing the light-matter interaction and inspire the exploration of nonlinear, non-Hermitian, and topological flat-band physics.

**II. Model Hamiltonian**

We start with a dipolar kagome lattice as shown in Fig. 1(a), in which each unit cell contains three lattice sites, and each site hosts a *p*-orbital (the two-colored dumbbell). The orientations of the three *p*-orbitals are initially set to be consistent with the '$q = 0$' frustrated spin configuration on a kagome lattice [40], and can be rotated simultaneously by an angle $\alpha$ relative to the original orientations, as shown in the lower-right inset of Fig. 1(a). Thus, the $C_3$ rotational symmetry is still preserved. Under the Wannier basis, the Hamiltonian of this dipolar kagome lattice is

$$H = \sum_{\langle i,j \rangle} t_{i,j} |i\rangle\langle j| + h.c., \qquad (1)$$

where $t_{i,j}$ is the hopping amplitude between two sites *i* and *j*. Different from that of the *s*-orbital, the interactions between *p*-orbitals are orientation-dependent. Here, we adopt the orientation dependence derived in Refs. [41-45] [see details in Supplementary Materials (SM) Sec. 1]:

$$t_{i,j} = \frac{1}{r_{ij}^3} (\cos\beta_i, \sin\beta_i) \begin{pmatrix} 3\cos^2\theta - 1 & 3\cos\theta\sin\theta \\ 3\cos\theta\sin\theta & 3\sin^2\theta - 1 \end{pmatrix} \begin{pmatrix} \cos\beta_j \\ \sin\beta_j \end{pmatrix}, \qquad (2)$$

where $r_{ij} = |\boldsymbol{r}_{ij}|$ with $\boldsymbol{r}_{ij} = \boldsymbol{r}_i - \boldsymbol{r}_j$ being the vector pointing from site *j* to site *i*, $\theta$ is the azimuthal angle of $\boldsymbol{r}_{ij}$, and $\beta_i$ and $\beta_j$ denote the corresponding orientation angles. The $1/r_{ij}^3$ term reflects the spatial decay of the interaction. In subsequent discussions, we first take nearest-neighbor (NN) approximation and set $1/r_{ij}^3 = 1$.

Figure 1(b) plots the band structures of the dipolar kagome lattice with a few typical rotation angles. Similar to the kagome lattice with *s*-orbitals, a flat band persists regardless of $\alpha$. In addition, there are two dispersive bands lie below the flat band at $\alpha = 0°$ (left panel) but above the flat band at $\alpha = 90°$ (right panel). The band structure at $\alpha = 180°$ is the same as that at $\alpha = 0°$ since the dipolar kagome lattice exhibits $C_2$ symmetry. This feature indicates that there should be a band flip process at some critical rotation angles (denoted as $\tilde{\alpha}$). To quantify this band evolution, we define the total bandwidth as

$$W \equiv \max_{\boldsymbol{k} \in \mathrm{BZ}} \left( E_3(\boldsymbol{k}) - E_1(\boldsymbol{k}) \right), \tag{3}$$

where BZ denotes the first Brillouin zone, and $E_3(\boldsymbol{k})$ and $E_1(\boldsymbol{k})$ represent the highest and lowest eigenenergies of the three bands. Figure 1(c) plots $\log(W)$ as a function of $\alpha \in [0, 180°]$, which shows clearly that $\log(W)$ drops to zero at $\alpha \in \{\tilde{\alpha}_1 = 49.8°, \tilde{\alpha}_2 = 130.2°\}$. $\log(W)$ is symmetric with respect to $\alpha = 90°$ since the two corresponding configurations of dipoles are related by the mirror symmetry. The middle panel of Fig. 1(b) shows the band structure of the system at $\alpha = \tilde{\alpha}_1$, in which the two dispersive bands are compressed to flat bands and degenerate with the original flat band. The band flip process and the emergence of fully flat bands at $\{\tilde{\alpha}_1 = 49.8°, \tilde{\alpha}_2 = 130.2°\}$ are closely related to the sign reversal of the dipole-dipole interaction during the rotation of $p$-orbital when

$$\tan^2(\tilde{\alpha} + \pi/6) - 3\sqrt{3} \tan(\tilde{\alpha} + \pi/6) - 2 = 0. \tag{4}$$

In a realistic physical system (e.g., PhC) where the orbitals are only approximately dipolar, the sign reversal in the effective dipole-dipole interaction persists, although the specific values of $\tilde{\alpha}$ may shift slightly (see details in SM Sec. 1). Similar to the $s$-orbital kagome lattice, the flat bands at $\tilde{\alpha}$ become dispersive when next-nearest-neighbor (NNN) couplings are introduced. However, under the same NNN coupling strength, the total bandwidth $W$ of all three bands in the dipolar kagome lattice is around one order of magnitude smaller than that of the single flat band in the $s$-orbital kagome lattice, demonstrating that our flat-band mechanism is significantly more resilient to NNN couplings (see details in SM Sec. 2).

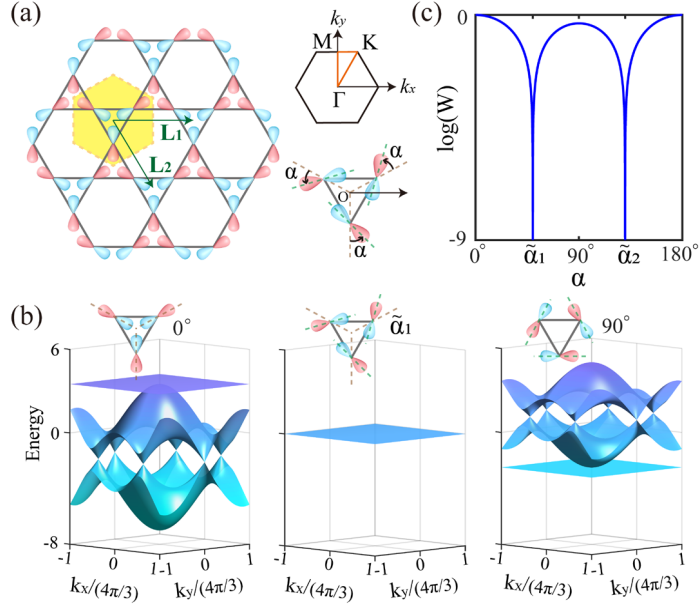

FIG. 1. Dipolar kagome lattice. (a) Lattice configuration (left panel) and the first Brillouin zone (upper-right panel). The yellow shaded region highlights a unit cell consisting of three $p$-orbitals, and $L_1$ and $L_2$ denote the primitive vectors. The three $p$-orbitals can be rotated simultaneously by an angle $\alpha$ relative to their original orientations (lower-right panel). (b) Band structures at $\alpha = 0°$ (left panel), $49.8°$ (middle panel) and $90°$ (right panel). The upper insets show the configuration of the dipole orientations within a unit cell. (c) Total bandwidth W of the three bands versus the rotational angle $\alpha$.

The presence of a flat band implies the existence of localized excitations, but in general only under appropriately designed source patterns. In such cases, the field profile faithfully replicates the source pattern, allowing for deterministic excitation of the system through tailored source distributions. To characterize the localization degree of the excited field, we define

$$P \equiv \langle S(\mathbf{r})|\varphi(\mathbf{r})\rangle, \qquad (5)$$

where $S(\mathbf{r})$ and $\varphi(\mathbf{r})$ denote normalized source and excited field distributions, respectively. Perfect localized excitation ($P = 1$) occurs when the $\varphi(\mathbf{r})$ exactly matches $S(\mathbf{r})$. Figure 2(a) illustrates the setup for characterizing $P$ in our system. We consider a finite but sufficiently large triangular system with the source placed at the center, which consists of six point-sources (dark red dots) with the same amplitude and a phase difference of $\delta$ between adjacent sources. To ensure all excited dispersive modes propagate only over a finite distance and to better match experimental conditions where material absorption is unavoidable, we introduce a small onsite absorption term $i\gamma$, with $\gamma = 0.3$ at each lattice site.

With the given source, the excited fields can be obtained using the Green's function method [46]. Here, the source energy is set to match the energy of the flat band. Without loss of generality, we first set $\delta = \pi/2$. Figure 2(b) plots $P$ as a function of $\alpha$ from 0° to 180°. $P$ reaches its maximum $P = 1$ at $\alpha = \{\tilde{\alpha}_1, \tilde{\alpha}_2\}$ where all three bands are flat, and drops when away from these two critical rotational angles (see the excited field distributions under different rotational angles in SM Sec. 3). Only the source points are excited when $\alpha = \{\tilde{\alpha}_1, \tilde{\alpha}_2\}$, whereas for other rotational angles, the excited fields spread out from the source position.

The $s$-orbital kagome lattice supports only a single flat band, such that localized excitation can be achieved only when $\delta = \pi$; otherwise, the dispersive bands will also be excited. Note that the excitation of dispersive bands under general source configurations is also unavoidable in other flat-band systems, such as the Lieb lattice [11,12]. In contrast, for the dipolar kagome lattice, localized excitation is independent of $\delta$ provided that $\alpha \in \{\tilde{\alpha}_1, \tilde{\alpha}_2\}$. Figure 2(c) plots $P$ as a function of $\delta$ for the $s$-orbital (red line) and the dipolar kagome lattices at $\tilde{\alpha}_1$ (blue line). $P$ keeps a constant value for the dipolar kagome lattice, while that of the $s$-orbital kagome lattice reaches its maximum only at $\delta = \pi$. For reference, the corresponding excited field distributions for a few typical $\delta$ are provided in SM Sec. 3. In fact, since all bands are flat for the dipolar kagome lattice at $\alpha = \{\tilde{\alpha}_1, \tilde{\alpha}_2\}$, perfect localized excitation is achieved under arbitrary source configurations (see details in SM Sec. 3 for a few other source patterns).

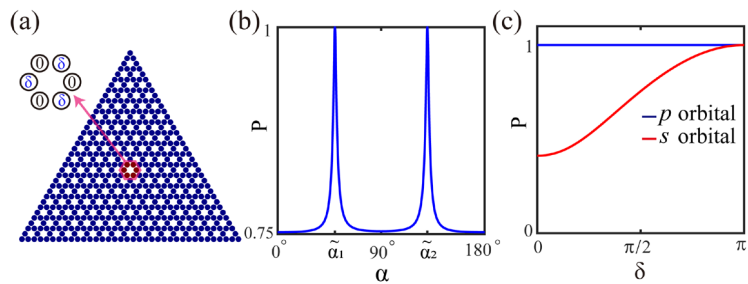

FIG. 2. Localized excitation under arbitrary source patterns. (a) Excitation scheme of the system, where the blue dots represent the lattice sites and dark red dots denote the source. Here we consider six point sources with the same amplitude but a phase difference $\delta$ between two adjacent sources (the upper left inset). (b, c) The localization degree of the excited fields as a function of $\alpha$ (b) and $\delta$ (c). The blue lines in (b-c) represent the dipolar kagome lattice, while the red line in (c) represents the $s$-orbital kagome lattice. We fix $\delta = \pi/2$ in (b) and $\alpha = 49.8°$ in (c).

**III. Photonic realization of the dipolar kagome lattice**

Now we implement the dipolar kagome lattice in a PhC, and the unit cell is illustrated in Fig. 3(a). The gray rectangular pillars represent dielectric materials with a relative permittivity of 25, the golden hexagonal pillar represents metal [modeled as perfect electric conductors (PEC) in simulations], and the remaining region is air. The lattice constant is $a = 18$mm. The length and width of the dielectric rectangular pillar are set as $l_1 = 7$mm, and $l_2 = 2.5$mm, respectively, such that the two $p$-modes oriented along the in-plane perpendicular directions are separated in frequency. The height of the dielectric and metallic pillars is $H = 6$mm, and an air gap with the height of 0.2mm is considered to fit the realistic experiment in Sec. IV. To minimize the effect of the NNN hopping, the metallic pillars are employed (see also discussions in SM Sec. 4). The edge length of the metallic pillars is $L$=4.5mm. The rotational degree of freedom ($\alpha$) is implemented by physically rotating the dielectric pillars.

Figures 3(b-d) show the typical eigenfield profiles of dipolar modes and higher-order modes [denoted by the dots in Fig. 3(e)]. The simulations are performed using COMSOL Multiphysics. As shown in Figs. 3(e-g), the photonic bands can be divided into different sets with each containing three bands, and there are three sets of bands of our interest as denoted by the shaded region in different colors. The modes below the red shaded region are s-orbital-like, thus not applicable to the theory in Sec. II. The first and second sets of bands have dipole-like modes, while the third exhibits a higher-order mode. As $\alpha$ increases from $0°$ to $90°$ (case for $90°$ to $180°$ is symmetric), the original flat band flips from the top to the bottom of the other two bands for the first (red-shaded) and second (green-shaded) sets of bands. For the higher-order mode (the third set of bands shaded in blue), the bandwidth also varies with the rotational angle as the hopping between the modes at different lattice sites also changes sign during the rotation. In stark contrast, there is no evident decrease in the bandwidth for the $s$-orbital-like modes. Figure 3(h) plots the full bandwidth of each set of bands for $\alpha \in [0, 90°]$. Note that the bandwidth minimum is reached at around $60°$, deviating from the theoretically predicted angle in Eq.

(4). This discrepancy arises primarily from two factors. First, the NNN coupling is ignored in the derivation of Eq. (4), while it is unavoidable in a realistic PhC. Second, the field distributions of modes near the dielectric pillars only approximately resemble the ideal dipole modes; consequently, the angular dependence of the hopping amplitude is close to—but not exactly—the form predicted by Eq. (2). Nevertheless, as long as the hopping amplitude changes sign with varying $\alpha$ and the NNN coupling remains weak, a minimum bandwidth is always expected near the point where the hopping flips sign. The field localization property of a finite PhC system is studied in SM Sec.5.

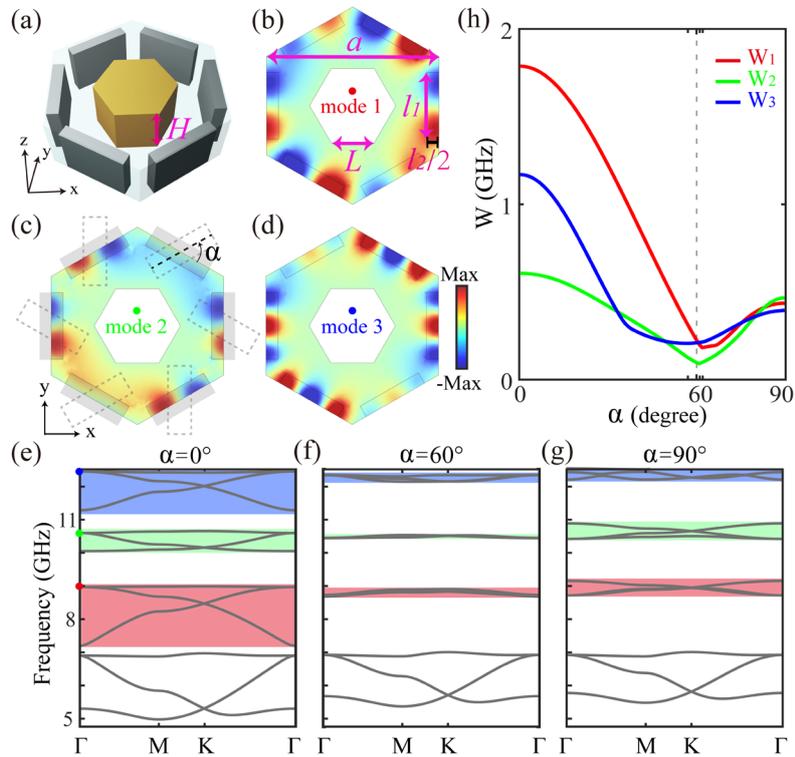

FIG. 3. Implementing dipolar kagome lattice in a PhC. (a) Schematic of the unit cell when the rotation angle is 0 degrees. (b-d) Typical mode profiles of three sets of bands (take $\alpha = 0°$ as examples). The first and second modes are dipole-like modes, while the third is a higher-order mode. (e-g) The band structures at three typical rotation angles $0°$, $60°$, and $90°$, respectively. (h) The variation of global bandwidth W of each set of bands. The minimum appears at round $60°$.

## IV. Experimental observation of the photonic flat bands

The PhC discussed above can be experimentally realized by sandwiching the entire structure between two parallel metallic plates. To experimentally observe the photonic flat bands, we fabricate three samples with rotation angles $\alpha = 0°$, $60°$, and $90°$, resepectively, as shown in Figs. 4(a-c). The lower-right insets show the zoomed-in views of the corresponding unit cells. For clarity, the top metallic plate is displaced to show the internal structure of the 2D PhCs. The gray dielectric rectangles and golden metallic hexagons are inserted in white foam to fix their positions. Unmagnetized yttrium iron garnet is used as the dielectric material with $\varepsilon = 25$, and all metallic components are fabricated with copper. We excite the 2D PhCs and measure their $E_z$ field distributions. The measured spatially resolved amplitude and phase distributions are Fourier transformed to obtain the band structures in reciprocal space, as shown in Figs. 4(d-f). Further details are provided in SM Sec.8. We can see that the measured results (colormaps) match well with simulated results (green dashed lines). Moreover, at $\alpha = 0°$ and $\alpha = 90°$, the flat bands are above and below the Dirac cone, respectively, and the three bands become relatively flat around 8.8 GHz and 10.5 GHz at $\alpha = 60°$. To demonstrate the field localization at the frequency of the flat bands, we excite the system with a single point current source [location marked by the cyan stars in Figs. 4(g-i)], and the measured $E_z$ distributions with different rotational angles at 10.55 GHz, as shown in Figs. 4(g–i). The fields are tightly localized at $\alpha = 60°$, whereas it spreads out at other two rotational angles.

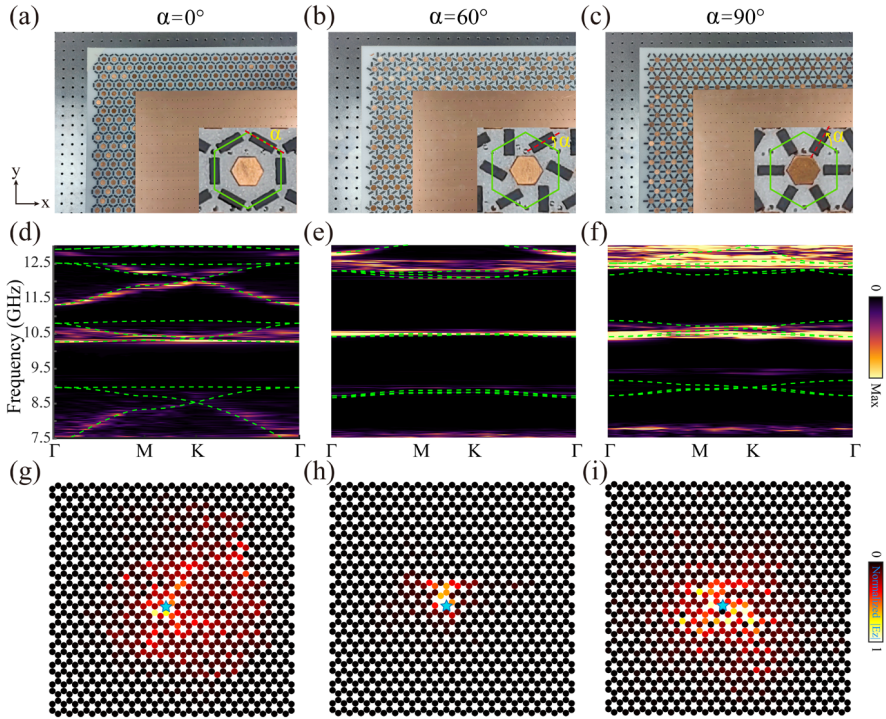

FIG. 4. Experimental observation of photonic flat bands. (a-c) Photograph of the three fabricated samples with $\alpha = 0°$, $60°$, and $90°$, respectively. The upper copper plate is shifted to show the inner structures. The insets are the zoom-in view of one unit cell (highlighted by a green hexagon). (d-f)

Measured (colormaps) and simulated (green dashed lines) band structures of the three 2D PhCs with different rotational angles. (g-i) Measured electric field distributions of the three 2D PhCs at 10.55 GHz. The cyan stars denote the single point source used for excitation. A second copper plate with hexagonally arranged probing holes is adopted to measure the field distributions (see details in SM Sec. 8).

## V. Summary

In conclusion, we have theoretically proposed and experimentally demonstrated a new mechanism for constructing multiply degenerate flat bands by utilizing *p*-orbital and rotational degrees of freedom. In contrast to the Lieb and *s*-orbital kagome lattices, the flat bands demonstrated here exhibit enhanced robustness against NNN couplings and do not require finely tuned source configurations for selective excitation. More significantly, our strategy is independent of specific lattice geometries or symmetries, and can be readily extended to other modes or lattices (see SM Sec. 6 and Sec. 7). Our work provides a broadly applicable platform for flat-band engineering and opens new avenues for controlling wave localization and interactions in photonic, electronic, and mechanical systems.

## Acknowledgement

Meng Xiao is supported by the Key Research and Development Program of the Ministry of Science and Technology (Grants No. 2024YFB2808200), the National Natural Science Foundation of China (Grant No.12321161645, Grant No. 12334015, and Grant No. 12274332), the National Key Research and Development Program of China (Grant No. 2022YFA1404900). Zhen Gao is supported by the National Natural Science Foundation of China under grants No. 62375118 and 62361166627, Guangdong Basic and Applied Basic Research Foundation under grant No.2024A1515012770, Shenzhen Science and Technology Innovation Commission under grants No. 20220815111105001 and 202308073000209, and High-level Special Funds under grant No. G03034K004.

**References**

[1] Y. Chen, J. Huang, K. Jiang, and J. Hu, Decoding flat bands from compact localized states, Sci. Bull. **68**, 3165 (2023).

[2] A. Nandy and A. Chakrabarti, Engineering flat electronic bands in quasiperiodic and fractal loop geometries, Phys. Lett. A **379**, 2876 (2015).

[3] Y. Cao *et al.*, Correlated insulator behaviour at half-filling in magic-angle graphene superlattices, Nature **556**, 80 (2018).

[4] X. Lu *et al.*, Superconductors, orbital magnets and correlated states in magic-angle bilayer graphene, Nature **574**, 653 (2019).

[5] E. Codecido *et al.*, Correlated insulating and superconducting states in twisted bilayer graphene below the magic angle, Sci. Adv. **5**, eaaw9770.

[6] Z. Liu, E. J. Bergholtz, H. Fan, and A. M. Läuchli, Fractional Chern Insulators in Topological Flat Bands with Higher Chern Number, Phys. Rev. Lett. **109**, 186805 (2012).

[7] E. J. Bergholtz and Z. Liu, Topological flat band models and fractional Chern insulators, Int. J. Mod. Phys. B **27**, 1330017 (2013).

[8] Y. Cao, V. Fatemi, S. Fang, K. Watanabe, T. Taniguchi, E. Kaxiras, and P. Jarillo-Herrero, Unconventional superconductivity in magic-angle graphene superlattices, Nature **556**, 43 (2018).

[9] M. Imada and M. Kohno, Superconductivity from Flat Dispersion Designed in Doped Mott Insulators, Phys. Rev. Lett. **84**, 143 (2000).

[10] R. Mondaini, G. G. Batrouni, and B. Grémaud, Pairing and superconductivity in the flat band: Creutz lattice, Phys. Rev. B **98**, 155142 (2018).

[11] R. A. Vicencio, C. Cantillano, L. Morales-Inostroza, B. Real, C. Mejía-Cortés, S. Weimann, A. Szameit, and M. I. Molina, Observation of Localized States in Lieb Photonic Lattices, Phys. Rev. Lett. **114**, 245503 (2015).

[12] S. Mukherjee, A. Spracklen, D. Choudhury, N. Goldman, P. Öhberg, E. Andersson, and R. R. Thomson, Observation of a Localized Flat-Band State in a Photonic Lieb Lattice, Phys. Rev. Lett. **114**, 245504 (2015).

[13] C. Elias *et al.*, Flat Bands and Giant Light-Matter Interaction in Hexagonal Boron Nitride, Phys. Rev. Lett. **127**, 137401 (2021).

[14] N. Myoung, H. C. Park, A. Ramachandran, E. Lidorikis, and J.-W. Ryu, Flat-band localization and self-collimation of light in photonic crystals, Sci. Rep. **9**, 2862 (2019).

[15] S. Longhi, Photonic flat-band laser, Opt. Lett. **44**, 287 (2019).

[16] C. Danieli, A. Andreanov, D. Leykam, and S. Flach, Flat band fine-tuning and its photonic applications, Nanophotonics **13**, 3925 (2024).

[17] D. Green, L. Santos, and C. Chamon, Isolated flat bands and spin-1 conical bands in two-dimensional lattices, Phys. Rev. B **82**, 075104 (2010).

[18] M. Tahir, O. Pinaud, and H. Chen, Emergent flat band lattices in spatially periodic magnetic fields, Phys. Rev. B **102**, 035425 (2020).

[19] E. Andrade, F. López-Urías, and G. G. Naumis, Topological origin of flat bands as pseudo-Landau levels in uniaxial strained graphene nanoribbons and induced magnetic ordering due to electron-electron interactions, Phys. Rev. B **107**, 235143 (2023).

[20] X. Wen, C. Qiu, Y. Qi, L. Ye, M. Ke, F. Zhang, and Z. Liu, Acoustic Landau quantization and quantum-Hall-like edge states, Nat. Phys. **15**, 352 (2019).

[21] M. C. Rechtsman, J. M. Zeuner, A. Tünnermann, S. Nolte, M. Segev, and A. Szameit, Strain-

induced pseudomagnetic field and photonic Landau levels in dielectric structures, Nat. Photonics **7**, 153 (2013).

[22] H. Abbaszadeh, A. Souslov, J. Paulose, H. Schomerus, and V. Vitelli, Sonic Landau Levels and Synthetic Gauge Fields in Mechanical Metamaterials, Phys. Rev. Lett. **119**, 195502 (2017).

[23] J. Yang, Y. Li, Y. Yang, X. Xie, Z. Zhang, J. Yuan, H. Cai, D.-W. Wang, and F. Gao, Realization of all-band-flat photonic lattices, Nat. Commun. **15**, 1484 (2024).

[24] O. Katz, G. Refael, and N. H. Lindner, Optically induced flat bands in twisted bilayer graphene, Phys. Rev. B **102**, 155123 (2020).

[25] Z. Zhang, Y. Wang, K. Watanabe, T. Taniguchi, K. Ueno, E. Tutuc, and B. J. LeRoy, Flat bands in twisted bilayer transition metal dichalcogenides, Nat. Phys. **16**, 1093 (2020).

[26] X. Zhang, T. Liu, Q. Zhang, X. Fan, F. Wu, and C. Qiu, Observation of ultraflat bands in gapped moire metamaterials, Phys. Rev. B **111**, 125143 (2025).

[27] Y. Li, H. A. Fertig, and B. Seradjeh, Floquet-engineered topological flat bands in irradiated twisted bilayer graphene, Phys. Rev. Research **2**, 043275 (2020).

[28] L. Du, X. Zhou, and G. A. Fiete, Quadratic band touching points and flat bands in two-dimensional topological Floquet systems, Phys. Rev. B **95**, 035136 (2017).

[29] H. Song and V. J. A. P. R. Van, All-Bands-Flat Floquet Topological Photonic Insulators with Microring Lattices, Adv. Photonics Res. **5**, 2400023 (2024).

[30] G. Li, L. Wang, R. Ye, S. Liu, Y. Zheng, L. Yuan, and X. Chen, Observation of flat-band and band transition in the synthetic space, Adv. Photonics **4**, 036002 (2022).

[31] D. Yu, G. Li, L. Wang, D. Leykam, L. Yuan, and X. Chen, Moire Lattice in One-Dimensional Synthetic Frequency Dimension, Phys. Rev. Lett. **130**, 143801 (2023).

[32] D. Weaire and M. F. Thorpe, Electronic Properties of an Amorphous Solid. I. A Simple Tight-Binding Theory, Phys. Rev. B **4**, 2508 (1971).

[33] C. Weeks and M. Franz, Topological insulators on the Lieb and perovskite lattices, Phys. Rev. B **82**, 085310 (2010).

[34] L. Morales-Inostroza and R. A. Vicencio, Simple method to construct flat-band lattices, Phys. Rev. A **94**, 043831 (2016).

[35] S. Mukherjee and R. R. Thomson, Observation of localized flat-band modes in a quasi-one-dimensional photonic rhombic lattice, Opt. Lett. **40**, 5443 (2015).

[36] F. Baboux *et al.*, Bosonic Condensation and Disorder-Induced Localization in a Flat Band, Phys. Rev. Lett. **116**, 066402 (2016).

[37] T. Zhang and G.-B. Jo, One-dimensional sawtooth and zigzag lattices for ultracold atoms, Sci. Rep. **5**, 16044 (2015).

[38] Y. Zong, S. Xia, L. Tang, D. Song, Y. Hu, Y. Pei, J. Su, Y. Li, and Z. Chen, Observation of localized flat-band states in Kagome photonic lattices, Opt. Express **24**, 8877 (2016).

[39] S. Taie, H. Ozawa, T. Ichinose, T. Nishio, S. Nakajima, and Y. Takahashi, Coherent driving and freezing of bosonic matter wave in an optical Lieb lattice, Sci. Adv. **1**, e1500854 (2015).

[40] L. Balents, Spin liquids in frustrated magnets, Nature **464**, 199 (2010).

[41] W. H. Weber and G. W. Ford, Propagation of optical excitations by dipolar interactions in metal nanoparticle chains, Phys. Rev. B **70**, 125429 (2004).

[42] L. Wang, R.-Y. Zhang, M. Xiao, D. Han, C. T. Chan, and W. J. Wen, The existence of topological edge states in honeycomb plasmonic lattices, New J. Phys. **18**, 103029 (2016).

[43] S. Y. Park and D. Stroud, Surface-plasmon dispersion relations in chains of metallic nanoparticles:

An exact quasistatic calculation, Phys. Rev. B **69**, 125418 (2004).

[44] Y.-R. Zhen, K. H. Fung, and C. T. Chan, Collective plasmonic modes in two-dimensional periodic arrays of metal nanoparticles, Phys. Rev. B **78**, 035419 (2008).

[45] K. H. Fung and C. T. Chan, Plasmonic modes in periodic metal nanoparticle chains: a direct dynamic eigenmode analysis, Opt. Lett. **32**, 973 (2007).

[46] T. Sondergaard and B. Tromborg, General theory for spontaneous emission in active dielectric microstructures: Example of a fiber amplifier, Phys. Rev. A **64**, 033812 (2001).